\documentclass[aps,prd,superscriptaddress,amsfonts,amssymb,amsmath,eqsecnum,nofootinbib,floatfix,twocolumn]{revtex4}
\usepackage{color,graphicx}
\usepackage{psfrag}
\usepackage[utf8]{inputenc}

\definecolor{darkblue}{rgb}{0,0,0.7}
\definecolor{darkred}{rgb}{0.7,0,0}
\usepackage[unicode, colorlinks, citecolor=darkblue, linkcolor=darkred, urlcolor=blue]{hyperref}
\usepackage{euler}

\allowdisplaybreaks

\begin{document}
\title{Back action cancellation in resolved sideband regime}
\author{Alena A. Demchenko and Sergey P. Vyatchanin}
\affiliation{Physics Department, Moscow State University, Moscow, Russia}
\date{\today}
\begin{abstract}
 We  consider the simple model of measurement  of mechanical oscillator position via Fabry-Pero cavity pumped by detuned laser (end mirror of cavity is mass of oscillator) in resolved sideband  regime when laser is detuned from cavity's frequency by  frequency of mechanical oscillator $\pm\omega_m$ and relaxation rate $\gamma$ of cavity is small: $\gamma\ll \omega_m$. We demonstrate fluctuation back action cancellation in reflected wave. However, it does not allow to circumvent Standard Quantum Limit, the reason of it is the dynamic back action.
\end{abstract}

\maketitle
\section{Introduction}
Vacuum fluctuations of mechanical oscillator displacement is a key prediction of quantum mechanics which is interesting to verify.  Optical parametric cooling of mechanical nano-oscillators close to their ground state \cite{VyDAN77, Kippenberg05, Wilson07, Kippenberg08, Marquardt09, OConnel10, Verlot09, Teufel09, Anetsberger09, Borkje10} makes it easier to observe the quantum behavior.
An impressive observation of a mesoscopic mechanical oscillator close to its ground state was recently made using the resolved sideband laser cooling \cite{Safavi12}.

The quantum fluctuations are responsible for quantum back action (disturbance of the quantum system) induced by a measuring device. As for continuous position measurement it results in an accuracy restriction (limitation) known as {\em Standard Quantum Limit} (SQL) first derived by Braginsky \cite{Braginsky68, BrKh92}.
Observation of back action as well as SQL is difficult to realize for mechanical system. But recently it became possible with the help of opto-mechanical devices that couple optical degrees of freedom to the mechanical oscillator  and that way approach the quantum regime \cite{Kippenberg08, Marquardt09, OConnel10}. Now several groups are close to this goal \cite{Verlot09,Teufel09,Anetsberger09,Borkje10,Westphal11}.

Usually the resolved sideband regime is used under the conditions of a frequency shift between the laser frequency and the frequency of a cavity mode. This shift equals to the frequency of the mechanical oscillator which is much larger than the optical bandwidth of the cavity mode \eqref{cond}. In this paper we analyze this regime in oder to find minimal signal force acting on a mechanical oscillator.  We show that  back action is canceled. However, it does not allow to circumvent SQL due to {\em dynamical} back action (introduction of damping into mechanical oscillator).

\section{Simplified model}

\begin{figure}[b]
\includegraphics[width=0.4\textwidth]{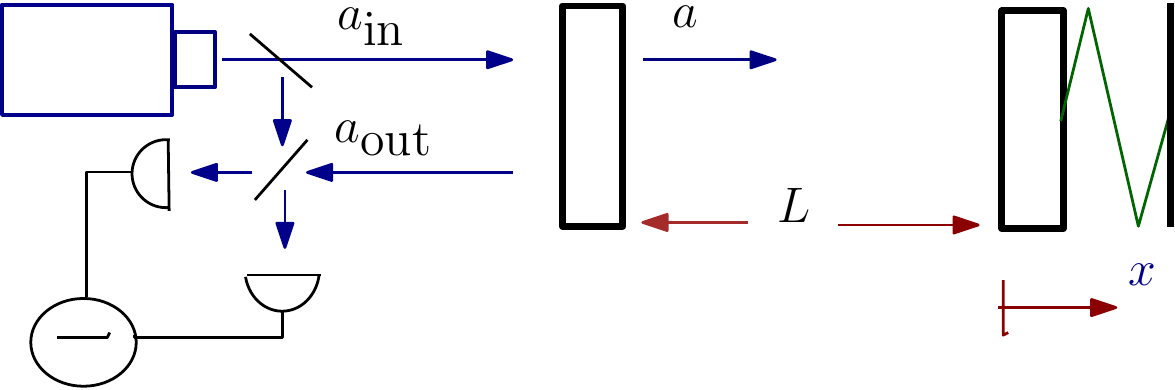}
\caption{Fabry-Perot cavity with one movable mirror.  Laser is detuned from cavity frequency. The measurement of quadrature in output wave provides information on mirror's displacement.} \label{resonP}
\end{figure}

 As the model of an opto-mechanical system we consider a 
Fabry-Perot cavity with a movable mirror (see Fig.~\ref{resonP}). (Note that this model is also valid for interaction of mechanical oscillator with light waves in toroidal microcavities \cite{Kippenberg05, Wilson07, Kippenberg08}.) The end mirror of the cavity acts as a mass of a mechanical oscillator of a frequency $\omega_m$ and a damping rate $\gamma_m$. The cavity is pumped by a laser of frequency $\omega_L$ detuned from resonant frequency $\omega_0$ of the cavity the mode by $\Delta =\omega_0-\omega_L$. In such opto-mechanical system back action is induced by the fluctuations of the light pressure force. Measuring the amplitude quadrature of the output wave one gets the information about the mechanical displacement. What we are interested in is the the minimal force(acting on the mechanical oscillator)that could be measured.

All further considerations are made under the condition of the resolved sideband regime:

\begin{equation}
 \label{cond}
 \gamma_m \ll \gamma \ll \omega_m,
\end{equation}
where  $\gamma$ is relaxation rate of the cavity mode.

The theoretical background for this model is well known \cite{Wilson07, Genes08, Borkje10, Miao10}. We start from simplified model considering two cases of positive and negative detuning ($\Delta = \pm \omega_m$)  separately.

\subsection{Positive detuning}
Let us consider the case of positive detuning $\Delta= \omega_m$ (laser frequency is {\em smaller} than cavity frequency).
There is a conventional set of equations in frequency domain for the wave amplitude $a(\Omega)$ inside the cavity and the Fourier
transform $b(\Omega)$ of the mechanical oscillator's annihilation operator \cite{Miao10, collettt_1984_parametric_amplification,
woolley_2008_microwave_cavity, 12a1KiMiYaSaPaChPRA}:
\begin{subequations}
\label{setP}
\begin{align}
   \label{setPa}
  	(\gamma  -i\nu)\, a + iG_0 b  &= \sqrt{2\gamma}\, a_\text{in},\\
   \label{setPb}  
 i G_0^*\, a + (\gamma_m - i \nu)\, b & =  	\sqrt{2\gamma_m}\, b_\text{th} +f,\\
   \label{setPc}
   a^\dag_- \big[\gamma - i(2\omega_m + \nu)\big] & = 
   	\sqrt{2\gamma}\, a^\dag_{\text{in}-},\\
	a_\text{out}   & = - a_\text{in}  + \sqrt{2\gamma}\, a,\\
	a\equiv a(\Omega), \quad a^\dag_{\text{in}-} & \equiv a^\dag_\text{in}(-\Omega),
		\quad 	\Omega  \equiv  \omega_m +\nu,\nonumber\\ 
	b\equiv b(\Omega), \quad b_\text{th} & \equiv b_\text{th}(\Omega),\quad
		a^\dag_-  \equiv a^\dag(-\Omega)\\
		\label{G0}
 |G_0| = \sqrt\frac{kI_0}{m L\omega_m},\quad 
 	&f = \frac{F(\Omega)}{i\sqrt{\hslash \omega_m m}}.
\end{align}
\end{subequations}
Here $a_\text{in}$ and $a_\text{out}$ describe the fluctuation amplitudes of the incident and reflected waves, $G_0$ is an opto-mechanical constant, $k$ is a wave vector of light wave, $I_0$ is the light power inside the  cavity, $m$ is the mass of the movable mirror, $L$ is a mean distance between the mirrors of the cavity, $F(\Omega)$ is a signal force (details in Appendix~\ref{app1}). 

The main simplification in the set \eqref{setP} is the equation \eqref{setPc} where  the interaction between the left sideband amplitudes $a^\dag(-\Omega)$ and the mechanical oscillator is omitted. This approximation is based on condition \eqref{cond}. 

We  calculate  determinant of the set (\ref{setPa}, \ref{setPb}):
\begin{align}
 D & 
    = (\gamma  -i\nu)\left(\gamma_m +\eta -i\nu\right),\\
    \label{eta}
   \eta & \equiv \frac{|G_0|^2}{\gamma-i\nu}= \eta_r +i\eta_i,\quad
   	\eta_r \equiv \frac{|G_0|^2\gamma }{\gamma^2 + \nu^2}\,.
\end{align}
Here $\eta$ describes ponderomotive rigidity which transforms into damping in resolved sideband case. Indeed, the real part $\eta_r$ is {\em positive} ponderomotive damping (it is this damping that is responsible for parametric cooling \cite{VyDAN77, Kippenberg05, 
Wilson07, Kippenberg08, Marquardt09, OConnel10, Verlot09, Teufel09, Anetsberger09, Borkje10}). Imaginary part $\eta_i$ is negligibly small ($\eta_i =\eta_r \nu/\gamma\ll \eta_r$) due to condition (\ref{cond}). One may easily find solution of (\ref{setP}) for $a(\Omega),\  a^\dag(-\Omega)$ and calculate $a_\text{out}$
\begin{align}
\label {aoutPos}
  a_\text{out}  & = \frac{\gamma + i\nu}{\gamma -i\nu}\left(1 - 
    	\frac{2\eta_r}{\gamma_m +\eta -i\nu) }\right)a_\text{in} -\\
     &\quad  - \frac{i G_0\sqrt{2\eta_r}}{|G_0|\big(\gamma_m +\eta -i\nu\big)}
	\sqrt\frac{\gamma+i\nu}{\gamma-i\nu}\, \left(\sqrt{2\gamma_m} b_\text{th} +f\right)\,, \nonumber\\ 
\label {aoutPos2}
a_{\text{out}-} & =
		\frac{\gamma - i(2\omega_m+\nu)}{\gamma + i(2\omega_m + \nu)} \, a_{\text{in} - }\,.
\end{align}
The second term in round brackets in (\ref{aoutPos}) ($\sim \eta_r$) describes back action  via light pressure force --- introduced damping $\eta_r$ is proportional to the power $I_0$ circulating in the cavity. However, it can be shown by straightforward calculation that back action  will be completely compensated.

Let us measure the quadrature $y= \big( a_\text{out}(\Omega) +a^\dag_\text{out}(-\Omega)\big)/\sqrt 2$ in the output wave. Then the double-sided spectral density $S_y$ of the output quadrature $y$ is as follows:
 \begin{align}
   \label{Spos}
   S_y (\Omega)&
     =\frac{1}{2}  + \frac{2 \eta_r\gamma_m\, n_T }{\left|\gamma_m +\eta -i\nu\right|^2},
 \end{align}
where $n_T$ is the mean number of the thermal photons in the mechanical oscillator. Formula (\ref{Spos}) is valid for positive frequencies, however, it is not a problem due to  $S_y(\Omega) = S_y(-\Omega)$. To calculate the spectral density $S_y$ we used conventional correlators (\ref{corr}).

Well known that the term describing back action noise in the output spectral density is proportional to the squared power $I_0$
circulating in the cavity --- see, for example, \cite{02a1KiLeMaThVyPRD, 12a1KiMiYaSaPaChPRA}. In our notations it corresponds to
the term proportional to $\sim \eta_r^2$ --- see definitions (\ref{G0}, \ref{eta}). However, we see that back action term $\sim
\eta_r^2$ is absent in \eqref{Spos}, hence, it  demonstrates compensation of back action in resolved sideband regime.

Note that the same result may be obtained for {\em any other} quadrature $y_\theta= \big( a_\text{out}(\Omega)\,e^{-i\theta} +a^\dag_\text{out} (-\Omega)\,e^{i\theta}\big)/\sqrt 2$ due to approximation \eqref{setPc} (which means that the left sideband $a^\dag(-\Omega)$ does not interact with the mechanical degree of freedom).
 
This simplified model is correct for the description of the mechanical cooling.  Indeed, straightforward calculation of the mechanical oscillator's mean energy gives 
\begin{align}
\label{Ep}
 \mathcal E_m & = 
    \hslash \omega_m \left(\frac{\gamma_m n_T}{\gamma_m + \eta_r} +\frac{1}{2}\right)
\end{align}
We see that the mean energy $\mathcal E_m$ decreases as the pump ($\eta_r$) increases --- it is a well known results \cite{VyDAN77, Kippenberg08,Marquardt09, OConnel10, Verlot09, Teufel09, Anetsberger09, Borkje10, Wilson07}.


\subsection{Negative detuning}

We consider the case of the negative detuning ($\Delta=-\omega_m$, laser frequency is {\em larger} than the cavity frequency) and start from the basic set of equations using the same notations as in \eqref{setP} (see details in Appendix~\ref{app1}):
\begin{subequations}
\label{setN}
\begin{align}
\label{setNa}
  	(\gamma  +i\nu)\, a_- +iG_0 b^\dag  & =  \sqrt{2\gamma}\, a_{\text{in}-},\\
\label{setNb}
 	-i G_0^*\, a_- + (\gamma_m+i\nu)\, b^\dag  & = \sqrt{2\gamma_m}\, b_\text{th}^\dag + f,\\
\label{setNc} 	
	a_\text{in} \big[\gamma - i(2\omega_m + \nu)\big] &= \sqrt{2\gamma}\, a_\text{in},\\ 
\label{setNd}
	a_\text{out} (\pm\Omega) = - a_\text{in}(\pm\Omega)  &+ \sqrt{2\gamma}\, a(\pm\Omega)\,.
\end{align}
\end{subequations}
Again the first two equations in \eqref{setN} form the system of equations with the determinant:
\begin{align}
 D   & = (\gamma  + i\nu)\left(\gamma_m +\eta +i\nu\right),\\
 \eta &=\frac{-|G_0|^2}{\gamma+i\nu}= \eta_r +i\eta_i,\quad 
   	\eta_r =\frac{-|G_0|^2\gamma }{\gamma^2 + \nu^2}\,,
\end{align}
Here $\eta_r$ is a {\em negative} ponderomotive damping. We find $a_-$ and calculate $a_\text{out}$
\begin{align}
 \label {aoutNeg}
 	a_{\text{out}-} 
      & =  \frac{\gamma - i\nu}{\gamma + i\nu}\left(1 - \frac{2\eta_r\,}{\gamma_m +\eta + i\nu }\right)a_{\text{in}-} -\\
     &\quad - \frac{i G_0\sqrt{2|\eta_r|}}{|G_0|\left(\gamma_m +\eta +i\nu\right)} \,
     \left(\sqrt{2\gamma_m}\, b_\text{th}^\dag +f\right),\nonumber\\
    a_\text{out} & = \frac{\gamma + i(2\omega_m + \nu)}{\gamma - i(2\omega_m + \nu)}\,a_\text{in}\,.
\end{align}
The second term in round brackets in (\ref{aoutNeg}) describes back action via the light pressure force. Similarly to the previous case, the double-sided spectral density $S_y$ of the output amplitude quadrature $y$ equals to:
 \begin{align}
 \label{Sneg}
  S_y &=  \frac{1}{2}  + \frac{2 |\eta_r|\gamma_m\, \big(n_T +1\big)}{\left|\gamma_m +\eta -i\nu\right|^2}\,.
 \end{align}
So we see that back action terms $\sim \eta^2$ are completely compensated in \eqref{Sneg}.

The case $\Delta = - \omega_m$ corresponds to the negative damping ($\eta_r < 0$) and, hence, to the mechanical heating. That can be shown by the direct calculation of the mechanical oscillator's mean energy:
\begin{align}
\label{En}
 \mathcal E_m & =
    \hslash \omega_m \left(\frac{\gamma_m n_T}{\gamma_m - |\eta_r|} 
       + \frac{1}{2}\,\frac{\big(\gamma_m +|\eta_r|\big)}{\big(\gamma_m -|\eta_r|\big)}\right)
\end{align}
 As we see it is a well known result --- the mean energy $\mathcal E_m$ increases with the $|\eta_r|$  \cite{VyDAN77, Kippenberg08,Marquardt09, OConnel10, Verlot09, Teufel09, Anetsberger09, Borkje10, Wilson07} due to introduction of the negative damping.

\section{Discussion}

It is important that back action cancellation shown above does not provide the possibility to surpass SQL. 
To show it for the case of positive detuning we rewrite the spectral density $S_y$ (\ref{Spos})  so that it would have a
form of the dimensionless force $f_s$ defined in \eqref{setPb}:
\begin{align}
 S_f &= \frac{\big(\gamma_m +\eta_r\big)^2 + \nu ^2}{4\eta_r} +\gamma_m n_T
\end{align}
For the case of zero mechanical damping $\gamma_m = 0$\footnote{For negative detuning the particular case of zero mechanical damping has no sense as ponderomotive negative damping creates instability.} 
the detection condition of the resonant signal force $f_s=f_0\cos\omega_mt$ is as follows
\begin{align}
 f_0^2 > \int_{\Delta\Omega} 2S_f(\Omega)\,\frac{d\Omega}{2\pi} \simeq \left(\frac{\eta_r}{2} + \frac{\Delta\Omega^2}{6\eta_r}\right)
  \frac{\Delta\Omega}{2\pi}
\end{align}
Optimizing this formula  over $\eta_r$  and putting $\Delta\Omega/2\pi \simeq 1/\tau$ ($\tau$  is the time throughout which
the signal force acts and it is measurement time) we get the value of the minimal detectable force:
\begin{align}
\label{sql}
 f_0 > \xi \, \frac{1}{\tau},\quad \text{or }
 F_0 > \frac{\hslash f_0}{x_0}= \xi\, \frac{\sqrt{2\hslash m\omega_m}}{\tau}\, ,
\end{align}
where $\xi$ is a factor about $1$. Obviously, formula (\ref{sql}) describes SQL \cite{Braginsky68, BrKh92}.

So we may conclude that, despite the fact that {\em fluctuation} back action is completely compensated, the {\em dynamical} back action (which corresponds to the introduced damping $\eta_r$) is responsible for the SQL restriction.

For the case of non-zero damping $\gamma_m$ and narrow bandwidth ($\Delta\Omega < \gamma_m$ or $\gamma_m\tau>1$) we have the minimum of $S_f$ at $\eta_r=\gamma_m$ and the minimal force is equal to
\begin{align}
\label{Fth}
 F_0 > \sqrt\frac{4\hslash m\omega_m \gamma_m(n_T+1)}{\tau}\,.
\end{align}
These formulas for the minimal signal force coincide with the usual one \cite{Braginsky68,BrKh92}.
We see that even for the case $n_T=0$ restriction of thermal fluctuations does not vanish --- in contrast to the formula \eqref{Spos} where at $n_T=0$ the second term vanishes. It may also be explained by the {\em dynamical} back action. 

For the case of negative detuning the introduced damping $\eta_r$ is negative and in order to compensate possible instability feed back should be used. The detailed analysis shows that formulas  \eqref{sql} and \eqref{Fth} are still valid for the negative detuning case. 

We emphasize that we used resolved sideband condition \eqref{cond}, that allows to make  calculations simple and obvious.
However, we also made accurate self-consistent calculations of  spectral density $S_y$ of the output amplitude quadrature in
general using the set \eqref{setPlus2} in Appendix \ref{app1} taking into account the interaction of both sidebands with the
mechanical oscillator. We found that the back action cancellation takes place for the measurement of the {\em amplitude} quadrature in
the output wave. However, if one measures the {\em phase} quadrature in the output wave the spectral density of the homodyne current
contents back action terms ($\sim \eta_r^2$) but these terms are small enough. For example, for the case of positive detuning the
formula \eqref{Spos} will content an additional term,  which may be estimated as:
\begin{align}
 \eqref{Spos}:&\quad S_y^\text{add} \simeq \frac{2\eta_r^2 }{|\gamma_m+\eta-i\nu|^2} 
 	\times \frac{\gamma^2}{\omega_m^2}\,,
\end{align}
Obviously, the last multiplier is small due to the resolved sideband condition \eqref{cond}.

\appendix

\section{Details of model}\label{app1}

In this Appendix we derive the main formulas. The electric fields $E_\text{in}$ in the incident wave pump and the corresponding mean intensities $ J_\text{in}$ of light beam can be written as follows \cite{02a1KiLeMaThVyPRD}: 
\begin{align}
E_\text{in} & \simeq  \sqrt{\frac{2\pi \, \hslash \omega_{L}}{Sc} }\, e^{-i\omega_L t}\times\\
        &\quad \times \left( A_\text{in} +\int_{-\infty}^\infty
         a_\text{in}(\Omega)\, e^{-i\Omega t} \frac{d\Omega}{2\pi}\right)
         +  \big\{\text{h.c.} \big\},\nonumber\\
J_\text{in} & =  \hslash \omega_L|A_\text{in}|^2,\\
       \label{comma}
      &  \big[a_\text{in}(\Omega),\,a_\text{in}^\dag(\Omega')\big]=2\pi\,\delta(\Omega -\Omega'),\\
     \label{corr}
     & \big\langle  a_\text{in}(\Omega)\,a_\text{in}^\dag(\Omega')\big\rangle=2\pi\,\delta(\Omega -\Omega'),\\
     & \big\langle  a_\text{in}^\dag(\Omega)\,a_\text{in}(\Omega')\big\rangle= 0,\nonumber
\end{align}
where $S$ is the cross section of the light beam, $c$ is the velocity of light, $a$ and $a^\dag$ are annihilation and creation operators.

For amplitude $a_1$ inside the cavity ($a_1\equiv a_1(\Omega)$) and reflected amplitude $a_\text{out}$ we have usual formulas
\begin{subequations}
\label{inita1}
\begin{align}
a_1 & = \frac{\sqrt{\gamma/\tau}\, a_\text{in}}{\gamma-i(\Omega-\Delta)}
   +\frac{A_1\, 2ikx(\Omega)}{2\tau\big(\gamma-i(\Omega-\Delta)\big)}, \\
A_1 & =\frac{A_\text{in}\sqrt{\gamma/\tau}}{\gamma+i\Delta},\quad
      \gamma \equiv \frac{T^2}{4\tau},\quad \tau\equiv \frac L c,\,\\
\label{initb}
a_\text{out} & = - a_\text{in} + 2\sqrt{\gamma\tau}\,a_1= \nonumber\\
      & = -\,  \frac{\gamma+i(\Omega - \Delta)}{\gamma-i(\Omega - \Delta)}\, a_\text{in}
	+ \frac{i G x(\Omega)}{\gamma + i(\Delta - \Omega)},\\
  G & \equiv  2k A_1\sqrt{\gamma/\tau},\ |G|^2 = \frac{4kI_0\gamma}{\hslash L},\ I_0 = \hslash\omega_0 |A_1|^2.\nonumber
\end{align}
\end{subequations}
Writing an equation for the mechanical oscillator we take into account the
light pressure force $F_\text{pm}$ acting on the mechanical oscillator:  
\begin{subequations}
\label{inita1pm}
\begin{align}
\label{mechEq}
 m&\big(\omega_m^2 -2i\gamma_m \Omega -\Omega^2\big) x(\Omega)
 	= F_\text{pm} + F_\text{th} +F\,,\\
 \label{FpmR}
   F_\text{pm} &=  2\hslash k A_1^*  a_1(\Omega)  + 2\hslash k A_1  a_1^\dag(-\Omega).
\end{align}
\end{subequations}
Here $F_\text{pm}$ is a fluctuation light pressure force and $F_\text{th}$ is a thermal force.
We express displacement $x$ through the annihilation and creation operators $a_m,\ a_m^\dag$:
\begin{subequations}
\label{initbx}
\begin{align}
x(t) & = x_0\left(a_m(t) + a_m^\dag(t) \right),\quad 
    x_0  \equiv \sqrt\frac{\hslash}{2m\omega_m},\nonumber\\
 a_m(t) & = \int_0^\infty \left[a_m(\Omega) \, e^{-i\Omega t} \right]\frac{d\Omega}{2\pi},\\
a_m^\dag(t) & = \int_0^\infty \left[a_m^\dag(\Omega) \, e^{i\Omega t} \right]\frac{d\Omega}{2\pi},
\end{align}
\end{subequations}
and rewrite \eqref{mechEq} for the Fourier transform of an annihilation operator $a_m$ (below we denote $a_m\equiv a_m(\Omega)$ )
\begin{subequations}
\label{initbx2}
\begin{align}
 a_m  &\left(i[\omega_m - \Omega] +\gamma_m \right) =\\
    &=\sqrt{2\gamma_m}\,b_\text{th}
      -\frac{x_0}{i\hslash}\big( F_\text{pm} + F\big),\nonumber\\
   \langle & b_\text{th}^\dag(\Omega) b_\text{th}(\Omega')\rangle  =2\pi\,n_T\,\delta(\Omega-\Omega')\,,\\
   \label{commb}
   \big[ & b_\text{th}(\Omega),\, b^\dag_\text{th}(\Omega')\big]  =2\pi\,\delta(\Omega-\Omega')\,.
\end{align}
\end{subequations}
Here $b_\text{th}$ is an operator describing thermal forces. 

\paragraph*{The set of equations.}For $a_1,\ b_m,\ a_\text{out}$ ($a_1$ is the fluctuation amplitude of the wave inside the cavity) using (\ref{inita1}, \ref{initbx}) it is convenient to write  (here we assume $\Omega >0$ and denote $a_1\equiv a_1(\Omega),\ a_{1-}\equiv a_1(-\Omega)$)
\begin{subequations}
\label{setB12}
 \begin{align}
  \label{setB1a}
    a_1 &\big[\gamma-i(\Omega - \Delta)\big] - \frac{A_1\, 2ikx_0}{2\tau}\, b_m
      = \sqrt\frac{\gamma}{\tau}\, a_\text{in},\\
   \label{setB1b}
    a_{1-}^\dag& \big[\gamma-i(\Omega + \Delta)\big] + \frac{A_1^*\, 2ikx_0}{2\tau}\, b_m
      = \sqrt\frac{\gamma}{\tau}\, a^\dag_{\text{in}-},\\    
   - 2i&kx_0\big[A_1^* a_1 +A_1a_{1-}^\dag \big] + b_m\big[\gamma_m + i(\omega_m-\Omega)\big] = \nonumber\\
  \label{setB1c}
  &\qquad =	\sqrt{2\gamma_m}\, b_\text{th} +f_s,\\
  \label{out}
  &\quad a_\text{out}(\pm\Omega)  = - a_\text{in}(\pm \Omega) + 2\sqrt{\gamma\tau}\,a_1(\pm\Omega)\,.
 \end{align}
\end{subequations}
We introduce notations in order to rewrite this set in the conventional form 
 \begin{subequations}
\label{setPlus2}
 \begin{align} 
 \label{a1a}
  a \equiv & \sqrt{2\tau}\,a_1,\quad G_0 \equiv  \frac{-A_1\, 2kx_0}{\sqrt{2\tau}},\\
   |G_0| & =  \sqrt\frac{kI_0}{m L\omega_m},\quad  I_0  = \hslash \omega_0|A_1|^2,\\
   \label{setB1aa}
   & a \big[\gamma-i(\Omega-\Delta)\big] + iG_0\, b_m = \sqrt{2\gamma}\, a_\text{in}(\Omega),\\
    \label{setB1bb}
    & a_-^\dag  \big[\gamma-i(\Omega+\Delta)\big] - iG_0^*\, b_m = \sqrt{2\gamma}\, a^\dag_{\text{in}-},\\
  \label{setB1cc}
  & i\big[G_0^* a + G_0 a_-^\dag\big] + b_m\big[\gamma_m + i(\omega_m-\Omega)\big] = \\
  &\qquad 	=\sqrt{2\gamma_m}\, b_\text{th} +f_s.\nonumber\\
  &\quad a_\text{out}(\pm\Omega)  = - a_\text{in}(\pm \Omega) + 2\sqrt{\gamma\tau}\,a_1(\pm\Omega)\,.
 \end{align}
\end{subequations} 
For the case of positive detuning $\Delta =\omega_m$ we make an approximation using the condition \eqref{cond}: we omit the term $(-iG_0^*\, b_m) $ in the Eq. \eqref{setB1bb} and the term $(iG_0 a_-^\dag)$ in \eqref{setB1cc}. As a result we obtain the set \eqref{setP}

For the case of negative detuning $\Delta=-\omega_m$ we take the set that is complex conjugated to \eqref{setPlus2} and making similar approximations we get \eqref{setN}.

\end{document}